\documentclass[aps,prd,twocolumn,nofootinbib]{revtex4-1}
\usepackage[T1]{fontenc}
\usepackage[utf8]{inputenc}
\usepackage[english]{babel}
\usepackage{url}
\usepackage{mathrsfs}
\usepackage{amsfonts}
\usepackage{amsmath}
\usepackage{amssymb}
\usepackage{amsthm}
\usepackage{graphicx}

\usepackage{appendix}
\usepackage{bm}
\usepackage{booktabs}
\usepackage{braket}
\usepackage{caption}
\usepackage{enumerate}
\usepackage{enumitem}
\usepackage[Bjornstrup]{fncychap} 
\usepackage{fnpos} %
\usepackage[bottom,multiple]{footmisc} 
\usepackage{lipsum}	
\usepackage{mathtools} 
\usepackage[]{natbib}	
\usepackage{quoting}
\quotingsetup{font=small}
\usepackage{sidecap}
\usepackage{subfig}
\usepackage{textcomp}
\usepackage{tikz}
\usetikzlibrary{decorations.pathmorphing,calc}
\usepackage{titlesec}
\usepackage{tensor}
\usepackage{xfrac}

\usepackage{csquotes}
\usepackage{helvet}
\usepackage{listings}

\usepackage[bookmarks,bookmarksopen=true,colorlinks,citecolor=blue,linkcolor=blue,urlcolor=blue]{hyperref}

\makeFNbelow
\newcommand{\be}{\begin{equation}}
\newcommand{\ee}{\end{equation}}
\newcommand{\bn}{\begin{equation*}}
\newcommand{\en}{\end{equation*}}

\newcommand{\cit}{\citep}		

\newcommand{\lie}{\pounds} 
\begin{document}

\rmfamily

\title{Smarr Formula for Lovelock Black Holes: a Lagrangian approach}
\date{\today}
\author{Stefano Liberati}
\author{Costantino Pacilio}
\affiliation{SISSA, Via Bonomea 265, 34136 Trieste, Italy}
\affiliation{INFN, Sezione di Trieste, Trieste, Italy}
\begin{abstract}
We argue that the Smarr Formula for black holes can be expressed in terms of a Noether charge surface integral plus a suitable volume integral, for any gravitational theory. The integrals can be constructed as an application of Wald's formalism. We apply this formalism to compute the mass and the Smarr Formula for static Lovelock black holes. Finally, we propose a new prescription for Wald's entropy in the case of Lovelock black holes, which takes into account topological contributions to the entropy functional.
\end{abstract}
\maketitle
\section{Introduction}
\label{sec:intro}
The Smarr Formula (SF) expresses the mass of a black hole in terms of its geometrical and dynamical parameters (angular momentum, electromagnetic potential, area, etc.) and it was first derived in the context of General Relativity. For vacuum GR solutions the SF has a geometrical interpretation: it is equivalent to the Komar integral, a boundary surface integral of the covariant derivative of a Killing vector field. The question naturally raises if it is possible to find a similar geometrical interpretation of the SF for arbitrary theories of gravitation, i.e. a generalization of Komar's construction to other theories than GR.

Progresses in this direction have been done in the recent years: in particular Kastor et al.\cit{kastor1} have shown that, for the particular class of Lovelock theories, it is indeed possible to construct a surface integral generalizing the Komar one. However, the special features of the Lovelock Lagrangian play a key role, and their method doesn't seem extendible to more general theories.

If one weakens the requirement that Komar integral be a pure surface integral, and allows for volume integral contributions, then a strong result holds: any diffeomorphism invariant theory of gravity admits a geometrical identity- we call it the Smarr Identity (SI)- which reduces to the Komar one in the GR case. In this letter we suggest that the SI gives always the correct Smarr Formula, and we use it to derive a SF for Lovelock theories.

We proceed as follows: in Section \ref{sec:proof} we review Wald's derivation of BH entropy and derive the Smarr Identity; in Section \ref{sec:examples} we compute the mass and the SF for static vacuum black hole solutions in Lovelock theories; in Section \ref{sec:topology} we discuss the role of topological contributions to the Smarr Formula; Section \ref{sec:remarks} contains an overview of the results with concluding remarks.

\section{The Smarr Identity}
\label{sec:proof}
In \citep{waldentropy1} the first law of black hole mechanics is derived for diffeomorfism invariant theories, by making use of the conserved Noether current associated to a special vector field. The BH entropy is then identified as a geometric functional of the Noether potential: this is the main result of Wald's construction, that we briefly review. 

Some comments are in order. The derivation makes certain non-trivial assumptions on the spacetime geometry: in particular one starts with a stationary spacetime with an internal boundary, identified with the future event horizon of a single black hole. In GR it is proved that the event horizon of a stationary BH is a Killing horizon generated by a Killing field $\xi^a$; although there is no generalization of the proof to higher curvature theories, all the known solutions are of this kind. Therefore we restrict our attention to black holes whose event horizon is Killing, and we assume that its generators can be regularly extended in both directions. Hypersurface orthogonality ensures that $\xi^a$ is tangent to non-affinely parametrized geodesics, whose inaffinity $\kappa$ is defined by
\be
\xi^a\nabla_a\xi_b=\kappa\xi_b.
\ee
If $\kappa\neq 0$ one can show that: (i) $\kappa$ is constant over the horizon; (ii) the horizon contains a spacelike $(D-2)$-dimensional surface where $\xi^a$ vanishes, called the "bifurcation surface" B. In the following we will assume this to be always the case. 

Apart from time translations, the spacetime will admit other possible spatial symmetries: we specialize to the case of rotational symmetries generated by a set of vector fields $\{\psi^a_i\}$ collectively denoted by $\vec{\psi}^a$. The Killing field $\xi^a$ can then be expressed as
\be
\label{eq:killing}
\xi^a=t^a+\vec{\Omega}\cdot\vec{\psi}^a
\ee
where $\Omega_i$ is called the "angular velocity" of the horizon around the i-th axis.

Given this preliminary setup, let us review Wald's derivation. Consider a collection of dynamical fields in $D$ spacetime dimensions, collectively denoted by $\phi$, including a metric tensor $g_{ab}$ plus other possible matter fields, whose dynamics is determined by a Lagrangian $D$-form $\mathbb{L}=\mathcal{L}\epsilon$, with $\epsilon$ the spacetime volume element. 

Under a generic variation $\delta\phi$ of the fields, the variation of $\mathbb{L}$  can be expressed as a sum of a bulk term plus a boundary one:
\be
\label{eq:varl}
\delta\mathbb{L}=\mathbb{E}_{\phi}\delta\phi+d\Theta(\phi,\delta\phi)
\ee
where the $(D-1)$-form $\Theta$ is locally constructed out of $\phi$ and $\delta\phi$. From \eqref{eq:varl} we read that the e.o.m. are $\mathbb{E}_{\phi}\doteq 0$ for each $\phi$\footnote{From now on the dot indicates equalities holding on shell.}.

In particular one can consider infinitesimal variations along a vector field $\xi$, $\delta\phi=\lie_\xi\phi$. By diffeomorphism invariance, to any vector field $\xi$ corresponds a Nother current $(D-1)$-form
\be
\label{eq:j}
\mathbb{J}[\xi]=\Theta(\phi,\lie_\xi\phi)-\xi\cdot\mathbb{L}
\ee
which is conserved on shell:
\be
d\mathbb{J}[\xi]=d\Theta(\phi,\lie_\xi\phi)-d\xi\cdot\mathbb{L}=-\mathbb{E}_\phi\lie_\xi\phi\doteq 0.
\ee
The conservation of $\mathbb{J}$ implies the existence of a $(D-2)$-form $\mathbb{Q}[\xi]$ \citep{wald1990}
\be
\mathbb{J}[\xi]\doteq d\mathbb{Q}[\xi]
\ee
called the "Noether potential" associated to $\xi$. 

$\mathbb{Q}$ enters in the definition of the conserved charges: indeed the Hamiltonian variation, associated with the flow of $\xi$, over an initial value surface $\Sigma$ with boundary $\partial\Sigma$, is given by \citep{waldentropy1}
\be
\label{eq:deltah}
\delta H[\xi]=\int_{\partial\Sigma}\bigl[\delta\mathbb{Q}[\xi]-\xi\cdot\Theta(\phi,\delta\phi)\bigr];
\ee
it is then natural to identify the variations of the energy $E$ and the angular momentum $\vec{J}$ at infinity as\footnote{$\delta E$ contains also work term contributions from long range fields, such as gauge fields.}
\begin{align}
&\delta E=\int_{S_\infty}\bigl[\delta\mathbb{Q}[t]-t\cdot\Theta(\phi,\delta\phi)\bigr],\\
&\vec{\delta J}=-\int_{S_\infty}\bigl[\delta\mathbb{Q}[\vec{\psi}]-\vec{\psi}\cdot\Theta(\phi,\delta\phi)\bigr]=-\int_{S_\infty}\delta\mathbb{Q}[\vec{\psi}] \label{eq:delj},
\end{align}
where $S_\infty$ is the outer boundary of $\partial\Sigma$, and the last equality of \eqref{eq:delj} follows from the fact that $\vec{\psi}$\, is tangential to $S_\infty$. (Notice that, as usual, the angular charges are defined up to a conventional minus sign.). If there is a $D-2$ form $\mathbb{B}(\phi)$ such that $\int\xi\cdot\Theta(\phi,\delta\phi)=\delta\int\xi\cdot\mathbb{B}(\phi)$, one defines the conserved Hamiltonian charge as
\be
H[\xi]=\int_{S_\infty}\mathbb{Q}[\xi]-\xi\cdot\mathbb{B};
\ee
in particular the angular momentum is exactly the Noether charge at infinity, modulo a sign:
\be
\label{eq:J}
\vec{J}=-\int_{S_\infty}\mathbb{Q}[\vec{\psi}].
\ee
If the field $\xi$ is taken to be the Killing field \eqref{eq:killing} generating the horizon, then equation \eqref{eq:deltah} implies the first law of black hole mechanics: let (i) $\xi$ be a dynamical symmetry, meaning that $\lie_\xi\phi\overset{.}{=}0$ for all the $\phi$'s, and (ii) $\delta\phi$ be a variation of the dynamical fields around the BH solution, such that $\delta\phi$ solves the linearized e.o.m.; then $\delta H[\xi]\overset{.}{=}0$, from which it follows \citep{waldentropy1,waldentropy2}
\be
\label{eq:firstlaw}
\delta E\doteq \frac{\kappa}{2\pi}\delta{S}+\vec{\Omega}\cdot\vec{\delta J}
\ee
where $S$ is $2\pi/\kappa$ times the integral of $\mathbb{Q}$ over the bifurcation surface:
\be
\label{eq:entropy1}
S=\frac{2\pi}{\kappa}\int_{B}\mathbb{Q}[\xi],
\ee
and eq. \eqref{eq:firstlaw} is obtained by the vanishing of the integral \eqref{eq:deltah} over an initial value surface with boundary $\partial\Sigma=S_\infty\cup B$, with $B$ the bifurcation surface of the black hole.
Since $\kappa/2\pi$ is the Hawking temperature, one interpets $S$ as the thermodynamical entropy of the BH\footnote{Note however that this identification fails if the dynamical fields have divergent components at the bifurcation surface. This circumstance occours, for example, in the case of gauge fields, but one can see that in this case the divergences at the horizon can be gauged out by an appropriate gauge fixing, thus recovering the correct expression for the entropy.}. 

Finally it is worth noting that for a general gravitational Lagrangian eq.\eqref{eq:entropy1} can be expressed as \citep{waldentropy2}:
\be
\label{eq:entropy2}
S=-2\pi\int_B E^{abcd}_R\hat{\epsilon}_{ab}\hat{\epsilon}_{cd}\bar{\epsilon}\quad,\quad E_R^{abcd}=\frac{\delta\mathcal{L}}{\delta R_{abcd}}
\ee
where $\bar{\epsilon}$ is the area element of $B$ and $\hat{\epsilon}_{ab}$ is the binormal to $B$. 

As shown in \citep{jacobsonentropy} the integral \eqref{eq:entropy1} needs not to be evaluated at the bifurcation surface, since it gives the correct entropy on any other cross section of the horizon. The proof makes use of the fact that, being $\xi$ a dynamical symmetry, eq.\eqref{eq:j} becomes
\be
\label{eq:smarr0}
\mathbb{J}[\xi]+\xi\cdot\mathbb{L}\doteq 0,
\ee
provided $\Theta(\phi,\delta\phi)$ vanishes when $\delta\phi=0$. Indeed, the authors of \citep{waldentropy2} suggest an algoritm giving a preferred "canonical" $\Theta_0$, among all the possible $\Theta$'s, which is covariant, depends linearly on $\delta\phi$ and vanishes if $\delta\phi=0$. However the definition of Theta suffers of the ambiguity associated to the freedom of adding a closed form $\Theta\rightarrow\Theta+d\alpha$ which in principle can spoil the above properties: we follow the authors of \citep{jacobsonentropy,waldentropy2} and restrict only to those $\alpha$'s preserving the mentioned properties of $\Theta$. Eq. \eqref{eq:smarr0} is then ensured. Integration over $\Sigma$ then gives
\be
\label{eq:smarr}
\oint_{\partial\Sigma}\mathbb{Q}[\xi]+\int_\Sigma\xi\cdot\mathbb{L}\doteq 0.
\ee
By linearity of $\mathbb{Q}$ w.r.t. $\xi$, using eq.s \eqref{eq:J} and \eqref{eq:entropy1}, we obtain
\be
\label{eq:smarr2}
\oint_{S_\infty}\mathbb{Q}[t]\overset{.}{=}TS+\vec{\Omega}\cdot\vec{J}-\int_\Sigma \xi\cdot\mathbb{L}\quad\text{(Smarr Identity)}
\ee
where we used $\partial\Sigma=S_\infty\cup B$. This is the Smarr Identity: in the next Section we implement it to derive a generalized Smarr Formula for Lovelock theories.
\section{Smarr Formula from the Smarr Identity}
\label{sec:examples}
In the very simple example of 4-dimensional GR the Smarr Identity gives exactly the Komar integral
\be
\oint_{\partial\Sigma}\mathbb{Q}[\xi]\doteq 0
\ee
because the Einstein-Hilbert Lagrangian vanishes on shell. Therefore we propose here to regard the SI as a generalization of the Komar integral for general diffeo-invariant theories of gravitation: in particular we show how it provides a generalized Smarr Formula for the class of Lovelock theories. In \ref{subsec:lovelock} we review general features of Lovelock theories; in \ref{subsec:mass} we obtain a general expression for the mass of static spherically symmetric Lovelock black holes; finally in \ref{subsec:smarrlovelock} the desired Smarr Formula is obtained. 
\subsection{Lovelock theories}
\label{subsec:lovelock}
Lovelock theories generalize $\Lambda$-GR theory and are the most general vacuum second order gravity theories in higher dimensional spacetimes \citep{lovelock}. The peculiar structure of the Lagrangian makes them easier to deal with, if compared with more general higher curvature theories. The Lagrangian in $D$ dimensions is
\be
\label{eq:lovelockl}
\begin{split}
&\mathbb{L}=\mathcal{L}\epsilon=\sum_{k=0}^m c_k\mathcal{L}^{(k)}\epsilon,\\
&\mathcal{L}^{(k)}=\frac{1}{2^k}\tensor*{\delta}{*^{a_1b_1}_{c_1d_1}^{\dots}_{\dots}^{a_kb_k}_{c_kd_k}}\tensor*{R}{*_{a_1}^{c_1}_{b_1}^{d_1}}\dots\tensor*{R}{*_{a_k}^{c_k}_{b_k}^{d_k}}
\end{split}
\ee
for generic constants $c_k$. Since for $m>\bigl[\frac{D}{2}\bigr]$ the antisymmetrized delta symbol vanishes, $m$ is restricted to be $m\le\bigl[\frac{D}{2}\bigr]$; moreover, if $m=\bigl[\frac{D}{2}\bigr]$, the integral of $\mathcal{L}^{(m)}$ is a topological invariant proportional to the Euler characteristic in $D$ dimensions, and therefore it doesn't contribute to the dynamics. The e.o.m. are
\begin{align}
\label{eq:eomlovelock}
&\sum_{k=0}^m\tensor*{\mathcal{R}}{*^{(k)}_{}^{r}_{s}}-\frac{1}{2}\delta^r_s\mathcal{L}\doteq 0,\\
&\tensor*{\mathcal{R}}{*^{(k)}_{}^{r}_{s}}=\frac{kc_k}{2^k}\tensor*{\delta}{*^{a_1}_{c_1}^{b_1}_{s}^{\dots}_{\dots}^{a_kb_k}_{c_kd_k}}\tensor*{R}{_{a_1}^{c_1}_{b_1}^r}\dots R_{a_kb_k}^{c_kd_k}.
\end{align}
Following the procedure descirbed in \citep{waldentropy2}, the "canonical"  $\Theta$ is
\be
\label{eq:thetalovelock}
\Theta_0(\phi,\delta\phi)=\sum_{k=0}^m\frac{kc_k}{2^{k-1}}\tensor*{\delta}{*^{a_1b_1}_{c_1d_1}^{\dots}_{\dots}^{a_kb_k}_{c_kd_k}}\nabla^{d_1}\delta g_{b_1}^{c_1}\dots \tensor*{R}{*_{a_k}^{c_k}_{b_k}^{d_k}}\epsilon_{a_1}
\ee
and the corresponding Noether charge is
\be
\label{eq:qlovelock}
\mathbb{Q}[\xi]=\sum_{k=0}^m\frac{kc_k}{2^{k-1}}\tensor*{\delta}{*^{a_1b_1}_{c_1d_1}^{\dots}_{\dots}^{a_kb_k}_{c_kd_k}}\nabla_{[a_1}\xi^{d_1]}\dots \tensor*{R}{*_{a_k}^{c_k}_{b_k}^{d_k}}\epsilon\indices{_{b_1}^{c_1}}
\ee
where the squared brackets indicate total antisymmetrization. Through eq.\eqref{eq:entropy2}, this gives the entropy of a Lovelock BH (\citep{jacobsonlovelock}, see also \citep{clunan}):
\be
\label{eq:entropy3}
S=\sum_{k=0}^m 4\pi kc_k\oint_B\underleftarrow{\mathcal{L}}^{(k-1)}\bar{\epsilon}
\ee
where the under-left arrow means that the object is evaluated w.r.t. the induced metric on $B$.
 
The Smarr Identity \eqref{eq:smarr2} reads 
\be
\label{eq:lovelocksmarr}
\oint_{S_\infty}\mathbb{Q}[t]\doteq TS+\vec{\Omega}\cdot\vec{J}-W.
\ee
Observe that the work term
\be
\label{eq:w}
W=\sum_{k=0}^{m}c_k\int_\Sigma\mathcal{L}^{(k)}\xi\cdot\epsilon
\ee
contains powers of the Riemann tensor up to degree $m$; one can however use the e.o.m. to lower the degree by one, thus reducing $W$ to an expression easier to work with: it is sufficient to trace \eqref{eq:eomlovelock} and solve for $\mathcal{L}^{(m)}$, the resulting $\mathcal{L}$ being
\be
\label{eq:reducedl}
\mathcal{L}\doteq \sum_{k=0}^{m-1}\biggl(\frac{2k-2m}{D-2m}\biggr)c_k\mathcal{L}^{(k)}.
\ee
Plugging this expression in $W$ we get the equivalent form
\be
\label{eq:reducedw}
W\doteq \sum_{k=0}^{m-1}\biggl(\frac{2k-2m}{D-2m}\biggr)c_k\int_\Sigma\mathcal{L}^{(k)}\xi\cdot\epsilon
\ee
for the work term. For example, the $\Lambda-GR$ Lagrangian in $D$ dimensions gives the Smarr Identity
\be
\oint_{\partial\Sigma}\nabla^a\xi^b\epsilon_{ab}+\frac{4\Lambda}{(D-2)}\int_\Sigma\xi^a\epsilon_a\doteq 0
\ee
in agreement with the results of \citep{kastor1,kastorads}.

So far we have been general. The main difficulties of eq.\eqref{eq:lovelocksmarr} are that (i) the integral of $\mathbb{Q}[\xi]$ is not yet expressed in terms of the mass $M$ of the BH, and (ii) the work term $W$ is a volume integral and therefore it requires the knowledge of the solution over the entire spacetime. These difficulties can be addressed under the additional hypothesis of staticity. As a preliminary, we derive a general expression for the mass of a static black hole in Lovelock theories.
\subsection{Mass of a Lovelock black hole}
\label{subsec:mass}
Consider a black hole solution. One is tempted to define the total mass as the ADM energy, namely the value of the Hamiltonian at spatial infinity $H[t]$. However in general the Hamiltonian at infinity receives divergent contributions from the maximally symmetric background. To regularize these divergences, one defines the total mass as $H[t]-H_0[t]$, where $H_0[t]$ is the ADM energy of the background metric. Thus we can use the expression \eqref{eq:deltah} for $\delta H$:
\be
\label{eq:m1}
M=\delta H[t]=\int_{S_\infty}\bigl[\delta\mathbb{Q}[t]-t\cdot\Theta(\phi,\delta\phi)\bigr].
\ee
We need to identify the asymptotic form of the line element: if we assume staticity, then the metric at infinity approaches a maximally symmetric background, i.e. Minkowski or (Anti-)deSitter. It is known \citep{wheelerlovelock} that static spherically symmetric BH solutions of Lovelock theory are all of the form
\be
\label{eq:static}
ds^2=-f(r)dt^2+\frac{dr^2}{f(r)}+r^{D-2}d\Omega^2_{D-2}.
\ee
For definiteness, we specify to the AdS case and keep $f(r)$ to scale as
\be
\label{eq:f}
f(r)=1+\frac{r^2}{l^2}-\frac{\mu}{r^{D-3}}+o(r^{-(D-3)}).
\ee
The Minkowski case is recovered in the limit $l\rightarrow\infty$.

Let us compute the two terms in \eqref{eq:m1} separately. For a metric of the form \eqref{eq:static} the integral of $\mathbb{Q}[t]$ simplifies drastically: from eq.\eqref{eq:qlovelock} one gets
\be
\label{eq:q1}
\oint_{S_\infty}\mathbb{Q}[t]=\lim_{r\rightarrow\infty}\sum_{k=0}^m\biggl[\frac{k\,c_k\,\gamma_k\,(1-f)^{k-1}\,f'}{r^{2k-2}}\biggr]r^{D-2}\Omega_{D-2}
\ee
where we defined $\gamma_k=(D-2)!/(D-2k)!$\,. Therefore
\be
\oint_{S_\infty}\delta\mathbb{Q}[t]=\lim_{r\rightarrow\infty}\sum_{k=0}^{m}\frac{k\gamma_kc_k}{r^{2k-D}}\frac{d}{dr}\biggl[(1-f)^{k-1}\delta f\biggr]\Omega_{D-2}.
\ee
This is a variation around the maximally symmetric background, so we have to take 
\be 
f(r)=1+\frac{r^2}{l^2}\qquad \delta f(r)=-\frac{\mu}{r^{D-3}} 
\ee
which yields
\be
\begin{split}
\oint_{S_\infty}\delta\mathbb{Q}[t]&=\sum_{k=0}^{m}(-1)^{k+1}\frac{k\gamma_k c_k (D-2k-1)}{l^{2k-2}}\mu\,\Omega_{D-2}\\
&=(\sigma-\gamma)\mu\,\Omega_{D-2},
\end{split}
\ee
where, for later convenience, we defined
\begin{align}
&\gamma=\sum_{k=0}^m (-1)^{k+1}\frac{k c_k (D-2)!}{l^{2k-2}(D-2k)!},\\
&\sigma=\sum_{k=0}^m (-1)^{k+1}\frac{k c_k (D-2)!}{l^{2k-2}(D-2k-1)!}.
\end{align}
The same way we compute the second piece of the l.h.s. of \eqref{eq:m1}:
\begin{widetext}
\begin{multline}
-\oint_{S_\infty}t\cdot\Theta=\lim_{r\rightarrow\infty}\sum_{k=0}^m \frac{k\gamma_k c_k (1-f)^{k-1}f}{r^{2k-2}}2 r^a\nabla_{[b}\delta g_{a]}^b r^{D-2}\Omega_{D-2}=\\
=\lim_{r\rightarrow\infty}\sum_{k=0}^m \frac{k\gamma_k c_k (1-f)^{k-1}}{r^{2k-2}}\biggl(-\frac{d\delta f}{dr}-\frac{(D-2)\delta f}{r}\biggr) r^{D-2}\Omega_{D-2}=\\
=\sum_{k=0}^m(-1)^{k+1}\frac{k c_k \gamma_k}{l^{2k-2}} \mu\, \Omega_{D-2}=\gamma \mu\, \Omega_{D-2}.
\end{multline}
\end{widetext}
Putting the two pieces together we get
\begin{equation}
\label{eq:m2}
M=\sigma \mu\, \Omega_{D-2}.
\end{equation}

Notice that this same expression had been already obtained in \citep{kastor3} by means of an Hamiltonian analysis. Our Lagrangian derivation agrees, and confirms that $H[t]$ is exactly the ADM energy.
\subsection{Smarr Formula for Lovelock black holes}
\label{subsec:smarrlovelock}
The expression \eqref{eq:m2} for the mass allows to rewrite the Smarr Identity \eqref{eq:lovelocksmarr} as a Smarr Formula, namely as an identity expressing the mass in terms of geometric  and dynamical parameters. It is sufficient to plug the asymptotic form of $f$, eq.\eqref{eq:f}, into eq.\eqref{eq:q1}. The result is
\begin{multline}
\label{eq:qt}
\oint_{S_\infty}\mathbb{Q}[t]=\lim_{r\rightarrow\infty}\sum_{k=1}^m\frac{(-1)^{k+1}k\,c_k\,\gamma_k}{l^{2k-2}}\Biggl[\frac{2r^{D-1}}{l^2}\\
+(D-2k-1)\mu\Biggr]\Omega_{D-2}.
\end{multline}
The first term in parentheses is divergent: this divergence can be regularized as we did for the BH mass, i.e. by subtracting the same integral evaluated w.r.t. the background Ads metric. This subtraction cancels the divergence exactly and one has
\be
\label{eq:gammam}
\oint_{S_\infty-\text{Ads}}\mathbb{Q}[t]=\biggl(1-\frac{\gamma}{\sigma}\biggr)M.
\ee
Thus, by adopting this regularization prescription, the Smarr Identity \eqref{eq:lovelocksmarr} becomes
\be
\label{eq:lovelocksmarr2}
\biggl(1-\frac{\gamma}{\sigma}\biggr)M\doteq TS-\hat{W}
\ee
where $\hat{W}$ is now the \emph{regularized} work term
\be
\hat{W}=\sum_{k=0}^{m}c_k\int_{\Sigma-\text{Ads}}\mathcal{L}^{(k)}\xi\cdot\epsilon
\ee
and $\vec{J}=0$ because of staticity.

Now we have to deal with the fact that the work term is a volume integral. As we anticipated, this constitutes a difficulty becauses it forces to know the solution on a whole hypersurface; by contrast, a surface integral would allow to specify only the asymptotic behaviours of the solution. However, in the case of static solutions under consideration, $W$ becomes a surface integral over $\partial\Sigma$: this follows from the fact that static solutions of Lovelock theories are all of the form \eqref{eq:static}. Substitution into $\mathcal{L}^{(k)}$ yields
\be
\label{eq:lk}
\mathcal{L}^{(k)}=\frac{\gamma_k}{r^{D-2}}\frac{d^2}{dr^2}\biggl[(1-f)^k r^{D-2k}\biggr]
\ee
and the regularized work term becomes 
\be
\begin{split}
&\hat{W}=\sum_{k=0}^{m}c_k\oint_{\partial\Sigma-\text{AdS}}W^{(k)}d\Omega_{D-2},\\
&W^{(k)}=\gamma_k\frac{d}{dr}\biggl[(1-f)^k r^{D-2k}\biggr],
\end{split}
\ee
which, as anticipated, is a surface integral. Therefore in Lovelock theories the generalized Smarr Formula \eqref{eq:lovelocksmarr2} holds for \emph{static} Lovelock black holes, where $\hat{W}$ is now a surface integral. It is interesting and insightful to compare \eqref{eq:lovelocksmarr2} with a similar but not identical expression obtained in \citep{kastor2, kastor3}: the authors there start from an Hamiltonian analysis and derive an extended first law with dynamical Lovelock couplings; integration of such a differential law produces the Smarr Formula. The two formulas can of course be shown to be equivalent.

In addition, notice that the expansion \eqref{eq:static}-\eqref{eq:f} of the metric at infinity still holds in the case of \emph{rotating asymptotically flat} black holes: therefore, by taking the limit of eq. \eqref{eq:gammam} for $l\to\infty$ we obtain the Smarr Formula
\be
\label{eq:rotatingsmarr}
\frac{(D-3)}{(D-2)}M\doteq TS-\vec{\Omega}\cdot\vec{J}-W
\ee
for such BHs, where no regularization for $W$ is needed in the asmptotically flat case; now, however, $W$ is not generically expressible as a surface integral.
\section{Topological work term}
\label{sec:topology}
As we observed, if $m=\frac{D}{2}$ the last term $\mathcal{L}^{(m)}$ of the sum \eqref{eq:lovelockl} is topological, and it doesn't contribute to the e.o.m.; nonetheless the Smarr Fromula \eqref{eq:lovelocksmarr2} receives contributions from it. This is evident already in the simple training case of the Einstein--Gauss--Bonnet theory of gravity in four dimensions: the Lagrangian of EGB theory is
\begin{align}
&\mathcal{L}=\frac{1}{16\pi G}(\mathcal{L}^{(1)}+\alpha\mathcal{L}^{(2)}),\\
&\mathcal{L}^{(1)}=R,\\
&\mathcal{L}^{(2)}=R^2-4R_{ab}R^{ab}+R_{abcd}R^{abcd}.
\end{align}
The second term $\mathcal{L}^{(2)}$ is topological in four dimensions, and therefore the BH solutions are the same as in vacuum GR; since they are Ricci flat, the Smarr Formula becomes
\begin{align}
& \frac{M}{2}\doteq TS-\Omega J-W, \label{eq:gbsmarr}\\
&W=\alpha\int_\Sigma K\xi\cdot\epsilon \label{eq:gbwork}
\end{align}
where $K$ is the Kretschmann invariant $R_{abcd}R^{abcd}$.

On the other hand, the Smarr Formula in vacuum GR is known to be
\be
\label{eq:grsmarr}
\frac{M}{2}\doteq T\frac{A}{4G}-\Omega J.
\ee
Now, Wald's entropy $S$ in \eqref{eq:gbsmarr} is not simply the Bekenstein entropy, but it receives a topological contribution $S^{\text{top}}$ from the Gauss--Bonnet part of the Lagrangian:
\be
\label{eq:entropygb}
\begin{split}
S&=\frac{A}{4G}+\frac{\alpha}{2G}\oint_B\underleftarrow{\mathcal{L}}^{(1)}\bar{\epsilon}\\
&=\frac{A}{4G}+\frac{2\pi\alpha}{G}\chi
\end{split}
\ee
where $\chi$ is the Euler characteristic of  the bifurcation surface. For a single BH $\chi=2$, and therefore by consistency the work term \eqref{eq:gbwork} must be equal to
\be
\label{eq:gbwork2}
W=\frac{2\alpha\kappa}{G}.
\ee
This is indeed the case (for example for the Schwarzschild solution the Kretschmann scalar is $K=48G^2M^2/r^6$ and, using $\kappa=1/2r_H$, eq.\eqref{eq:gbwork2} follows).

By generalizing the above argument, we can conclude that, if $m=\frac{D}{2}$, the Smarr Formula always contains suitable "topological"' terms, performing the task of compensating the topological correction to the entropy.

In the case of spherically symmetric solutions, it is very easy to verify explicitely how the compensation arises (see Appendix): indeed it turns out that the topological counterterms sum up to give the temperature $T=f'(r_H)/4\pi$, times a surface integral at the bifurcation surface, which reproduces exactly $S^{\text{top}}$. Thus the compensation occours between terms having the very same geometrical nature. 
 
\rm This fact suggests that $S^{\text{top}}$ and its counterterms are not genuine physical contributions, respectively, to the entropy and to the work terms, but they are rather an artefact of Wald's formalism.

Indeed, the topological correction to the Bekenstein entropy in four dimensions has been addressed by several authors \citep{jacobsonentropy,liko,wallviolation}, arguing that it can lead to possible violations of the generalized $2^{\text{nd}}$ law \footnote{See however \citep{effectivegb}, in which the authors argument that such a violation doesn't occour, if the Gauss--Bonnet term is viewed as an effective field theory contribution.}: this again suggests that the physical entropy should be identified with the Bekenstein one, rather than Wald's one. After all, it would be quite strange that a physical quantity like the entropy be affected by terms in the Lagrangian not contributing to the dynamics.

How does this reconcile with eq.\eqref{eq:entropygb}? One could simply remove by hand the topological term from Wald's entropy, as suggested in \citep{wallviolation}. However, having we interpreted $S^{\text{top}}$ as an artefact of the formalism, we wonder if there is a natural window inside the formalism itself: the answer is in the affirmative. One can make use of a further ambiguity in the definition of $\mathbb{Q}[\xi]$, in addition to those listed in \citep{jacobsonentropy,waldentropy2}: as noted in \citep{clunan}, it is possible to rescale the Noether form by a term proportional to the volume element $\underleftarrow{\epsilon}$ of $S^2$,
\be 
\mathbb{Q}[\xi]\rightarrow\mathbb{Q}[\xi]+\text{const.}\cdot\underleftarrow{\epsilon},
\ee
where $\underleftarrow{\epsilon}$ is defined as
\be
\underleftarrow{\epsilon}=\frac{1}{2}\sin\theta\, d\theta\wedge d\phi,
\ee
without affecting the validity of Wald's construction, because $d\underleftarrow{\epsilon}=0$, and $\oint_{B\cup S_\infty}\underleftarrow{\epsilon}=0$. 

Therefore in four dimensional EGB we can redefine
\be
\mathbb{Q}[\xi]\rightarrow\mathbb{Q}[\xi]-\frac{\kappa\alpha}{4\pi G}\chi\underleftarrow{\epsilon},
\ee
so that the modified Noether potential gives the correct physical entropy, i.e. the Bekenstein one. The procedure can be straightforwardly generalized to higher dimensions.  Observe that $\underleftarrow{\epsilon}$ is well defined also in the rotating case, and our prescription is thus completely general.
\section{Discussion}
\label{sec:remarks}
In this work we presented a general procedure to compute the Smarr Formula for black holes, in any diffeoinvariant theory of gravity. The method makes use of eq. \eqref{eq:smarr2}, which is obtained integrating and expanding eq. \eqref{eq:smarr0}. 

To the extent of our knowledge, the above eq.s have been considered before, but not in connection with the Smarr Formula: in particular, eq. \eqref{eq:smarr0} was used in \citep{jacobsonentropy} to show that Wald's entropy formula can be evaluated not only over the bifurcation surface, but over any spatial cross section of the horizon.

We applied our procedure to the case of Lovelock black holes, thus deriving the Smarr Fomulas \eqref{eq:lovelocksmarr2} for static black holes, and \eqref{eq:rotatingsmarr} for rotating asymptotically flat black holes. In particular, static BHs show the preferable feature that the work term $W$ is a surface integral, which follows from the simple form \eqref{eq:static} that the line element assumes in the static BH solutions of Lovelock gravity. The derivation cannot be straightforwardly extended to the rotating case, because there is no general form of the line element. It would be interesting to investigate under which restrictions the relative extension can be done.

In the final part of the paper, we examined the behaviour of topological terms in the Lovelock Lagrangian; we argued that the corresponding topological terms in the Smarr Formula, including the contribution $S^{\text{top}}$ to the entropy, can be viewed as unphysical artefacts of the formalism; motivated by this, we proposed a modified prescription for the Noether charge, which incorporates topological effects and reconciles the results with the physical quantities.
\begin{acknowledgements}
S.L. acknwoledges financial support from the John Templeton Foundation (JTF) grant \#51876.
C.P. is extremely grateful to A. Mohd for elucidating discussions.
S.L. and C.P. are extremely grateful to T. Jacobson for comments on an earlier version of the manuscript.
\end{acknowledgements}
\newcounter{chapter}
\section*{Appendix}
Consider the Lovelock Lagrangian in even dimension $D$ and with $m\equiv\frac{D}{2}$, such that $\mathcal{L}^{(m)}$ is topological and doesn't contribute to the e.o.m.; nevertheless the Smarr Formula \eqref{eq:lovelocksmarr2} contains three different topological contributions: the first is the topological entropy component 
\be
\label{eq:stop1}
S^{\text{top}}:=c_m\frac{d S}{d c_m}=4\pi mc_m\oint_B\underleftarrow{\mathcal{L}}^{(m-1)}\bar{\epsilon}.
\ee
Given that 
\be
\tensor*{\underleftarrow{R}}{_a^c_b^d}=\frac{\tensor*{\bar{\delta}}{_a^b_c^d}}{r_H^2},
\ee
where $\tensor*{\bar{\delta}}{_a^b_c^d}$ is the antisymmetrized delta on the $(D-2)$-dimensional bifurcation surface, $S^{\text{top}}$ becomes
\be
\label{eq:stop2}
S^{\text{top}}\equiv\frac{2\pi D!\,c_{D/2} \Omega_{D-2}}{(D-1)}.
\ee
The other two contributions, as anticipated before in Section \ref{sec:topology}, compensate exactly $TS^{\text{top}}$. Let us show how the compensation occours. The second contribution is the topological part of $\hat{W}$,
\begin{widetext}
\begin{multline}
\label{eq:wtop}
\hat{W}^{\text{top}}:=c_m\frac{d\hat{W}}{d c_m}=c_{D/2}\gamma_{D/2}\oint_{\partial\Sigma-AdS}\frac{d}{dr}(1-f)^{D/2}d\Omega_{D-2}\\
=\frac{D!\,c_{D/2}}{2(D-1)}\oint_B(1-f)^{\frac{D}{2}-1}f'(r)d\Omega_{D-2}-\frac{D!\,c_{D/2}}{2(D-1)}\oint_{S_\infty-AdS}(1-f)^{\frac{D}{2}-1}f'(r)d\Omega_{D-2}.
\end{multline}
\end{widetext}
Finally, the last contribution comes from the l.h.s. of \eqref{eq:lovelocksmarr2}:
\be
\label{eq:mtop}
M^{\text{top}}:=c_m\frac{d }{d c_m}\Bigl(\frac{\gamma}{\sigma}M\Bigr)=\frac{(-1)^{\frac{D}{2}-1}D!\,c_{D/2}\,\mu\,\Omega_{D-2}}{2(D-1)l^{D-2}}.
\ee
Using \eqref{eq:f}, a direct calculation shows that the second term in \eqref{eq:wtop} cancels exactly \eqref{eq:mtop}. 
Therefore $TS^{\text{top}}$ is ultimately compensated by the first term on the r.h.s of \eqref{eq:wtop}: this consists of a surface integral over the bifurcation surface $B$; moreover, using $f(r_H)=0$ and $T=f'(r_H)/4\pi$, it's immediate to see that it factorizes precisely as $T$ times $S^{\text{top}}$. 

This shows that the topological terms in the Smarr Formula compensate with the same geometrical structure. 
\bibliography{entropy,smarrformula,gbsolutions,others}
\end{document}